**Vibe synthesis: hierarchical agentic decision-making for intent-driven multistep molecule synthesis**

Jie Chen[1,2#], Guikun Chen[3,8#], Feng Wan[1,2#], Ning Xu[1,2#], Wen Liu[1,2], Yixiang Ruan[1,2], Yiyi Qiu[1,2], Haohua Li[1,2], Wei Wang[6], Yi Yang[6], Xiaoxue Wang[6], Peng Wang[6], Jiazhou Zhu[7], Jun Xuan[5*], Chengxi Li[1,2*], Jianhong Xu[4*], Wenguan Wang[3*], Yiming Mo[1,2*]

[1]College of Chemical and Biological Engineering, Zhejiang University, Hangzhou 310027, China.

[2]Key Laboratory of Intelligent Manufacturing for Functional Chemicals, ZJU-Hangzhou Global Scientific and Technological Innovation Center, Hangzhou 311215, China.

[3]The State Key Lab of Brain-Machine Intelligence, Zhejiang University, Hangzhou, 311121, China.

[4]State Key Lab of Chemical Engineering and Low-carbon Technology, Department of Chemical Engineering, Tsinghua University, Beijing 100084, China.

[5]Department of Chemistry, Zhejiang University, Hangzhou 310058, China.

[6]ChemLex Technology Co. Ltd., Shanghai 201210, China.

[7]Suzhou Novartis Technical Development Co. Ltd., Changshu 215537, China.

[8]Harvard Medical School, Boston 02115, USA.

#These authors contributed equally to this work.

*Corresponding authors: Yiming Mo (email: yimingmo@zju.edu.cn), Wenguan Wang (email: wenguanwang@zju.edu.cn), Jianhong Xu (email: xujianhong@tsinghua.edu.cn), Chengxi Li (email: cxli@zju.edu.cn), Jun Xuan (email: xuanjun2021@zju.edu.cn).

## Abstract

**The ability to construct complex molecules underpins advances in medicine and the molecular sciences. Yet long-horizon multistep synthesis requires decisions to be revised as evidence accumulates across chemically dependent operations, a capability not readily supported by artificial-intelligence tools and automated platforms organized around predefined tasks and hard-coded workflows. Here we propose and demonstrate a vibe synthesis system (VibeSyn), an intent-driven architecture in which hierarchical agents translate chemist-defined molecular objectives and accumulating evidence into adaptive decisions executed through computational tools and laboratory automation. We evaluate VibeSyn in three demonstrations of increasing scope and autonomy, progressing from evidence-grounded experimental decisions to adaptive execution of a ten-step total synthesis of catharanthine and structure-to-compound synthesis of cyclic peptides containing unnatural amino acids (uAAs). Across the long-horizon campaigns, active human input and simple confirmation account for less than 7.5% and 5.2%, respectively, of recorded workflow actions and outcomes. VibeSyn points towards a future in which chemists define synthetic intent and agent-directed laboratories autonomously navigate the evolving experimental trajectory required to deliver the target compound.**

## Introduction

The emergence of vibe coding represents a new paradigm of human–machine interaction, where users communicate high-level intent through natural language rather than directly writing codes, while large language model (LLM)-based agents translate these intentions into executable programs through iterative generate–execute–test–refine cycles (Fig. 1)[1,2]. This shift from implementation-driven to intent-driven interaction is transforming how complex coding tasks are approached. Molecular synthesis presents a comparable challenge: transforming a target structure into an isolated and validated compound requires iterative design–make–test–analyze (DMTA) cycles, in which experimental outcomes continuously influence subsequent decisions[3,4]. Yet synthesis remains highly dependent on expert decisions and labor-intensive laboratory execution, limiting the pace and scalability of molecular discovery across pharmaceuticals, agrochemicals, and functional materials[5]. Extending the intent-driven paradigm to molecular synthesis through "vibe synthesis" could enable chemists to define synthetic goals while artificial intelligence (AI) agents integrate chemical knowledge, experimental evidence, and computational tools to guide evolving synthesis decisions.

Achieving this vibe-synthesis vision requires integrating AI with laboratory automation to enable systems capable of planning and executing synthesis campaigns across multiple stages[6]. Recent advances have established essential components: computer-aided synthesis planning and generative models for proposing synthetic routes[7–9], predictive models for recommending reaction conditions[10,11], AI-assisted analytical tools for structure elucidation[12–14], and robotic platforms for automating reaction execution, work-up, purification, and characterization[15–19]. However, multistep synthesis cannot be reduced to a sequential execution of a predefined plan, because experimental evidence can invalidate previous assumptions and require decisions at different stages[20]. Such decisions span multiple levels including strategic route selection, tactical choices of reactions and procedures, and operational execution (Fig. 1a)[4]. Connecting existing AI tools and automated platforms typically relies on explicitly programmed workflows, making it difficult to incorporate new chemical knowledge or context-dependent judgments without additional coding. Human chemists therefore remain involved throughout synthesis, interpreting experimental evidence and translating their judgments into executable workflow changes (Fig. 1b).

LLM-based agents[21], the core enabling technology behind vibe coding, provide a potential decision layer for vibe synthesis (Fig. 1b). LLMs have rapidly progressed from chemical knowledge chatting[22] to literature retrieval[23], synthesis protocol generation[24], and multimodal interpretation[25,26], while tool-augmented agents further connect language-based reasoning with computational methods[27] and laboratory automation[28]. Recent advances extend these capabilities to enable closed-loop reaction optimization[29,30] and multi-agent systems for complex task execution, such as coordination of domain-specific agents for nanobody design in the Virtual Lab[31] and collaborative hypothesis generation and refinement in Co-Scientist[32] and Robin[33]. However, long-horizon multistep synthesis presents a distinct decision-making challenge[34–36]: integrating experimental evidence across chemically dependent steps to determine whether to proceed as planned, correct an operational failure, refine a procedure, or revisit an upstream route choice. Meeting this challenge requires campaign-level, cross-hierarchical decision-making that translates these judgments into coordinated changes across the remaining synthesis, connecting local experimental observations to the overall molecular objective.

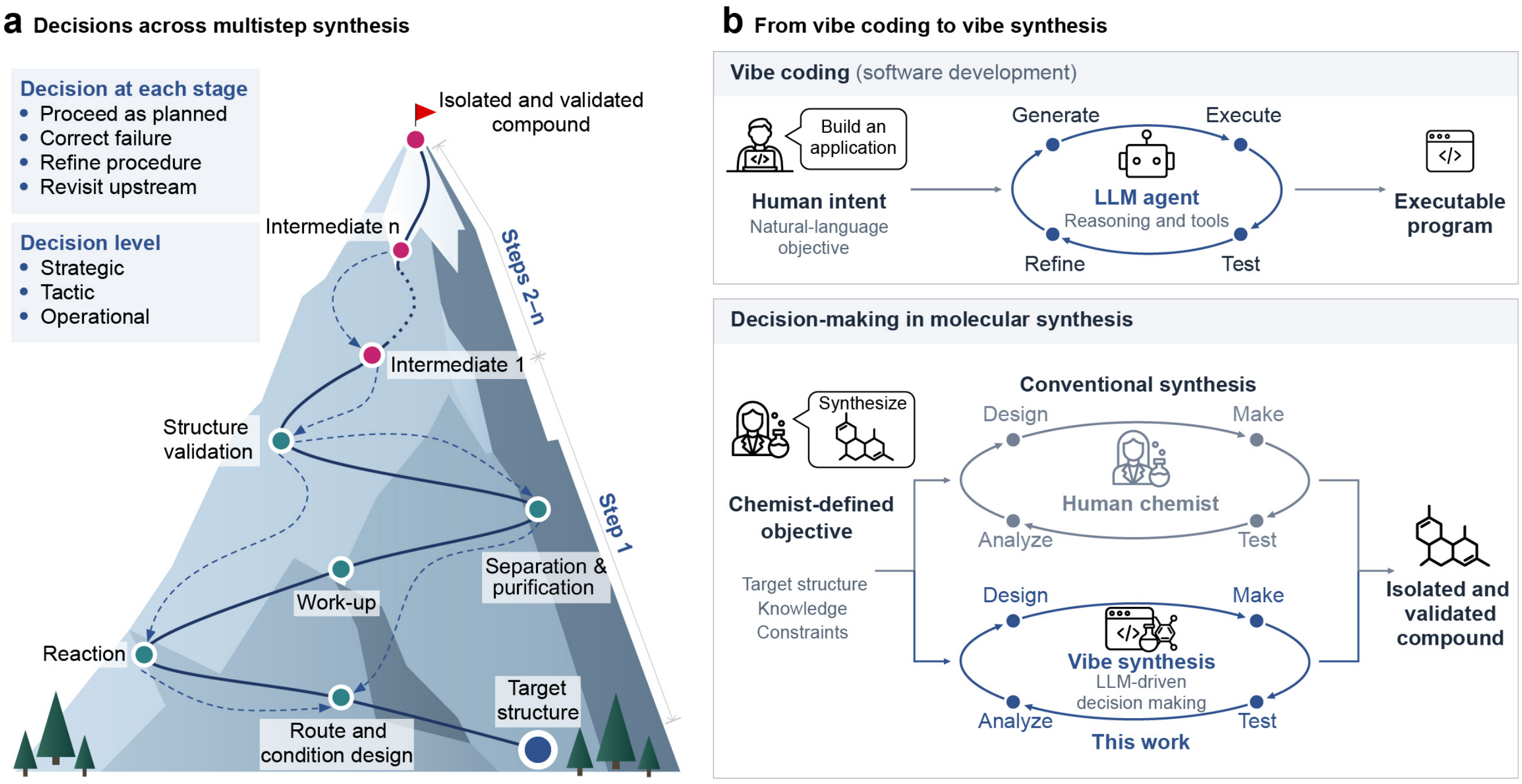

**Fig. 1 | Synthesis as an evidence-dependent decision process.** **a**, Experimental checkpoints and strategic, tactical, and operational decision levels across multistep synthesis. **b**, Comparison of vibe coding, conventional synthesis with programmed workflows, and vibe synthesis.

Here we introduce a vibe synthesis system (VibeSyn) that translates synthetic intent into adaptive decisions for automated molecular construction (Fig. 2a). A Manager Agent (MA) maintains the campaign-level objective and accumulated evidence, coordinating task agents that formulate inputs for computational tools, interpret tool outputs and multimodal experimental observations, and translate the resulting decisions into executable operations for automated synthesis devices. Human chemists define the objectives, constraints, and safety boundaries, and provide targeted guidance or confirmation when operations exceed validated chemical or hardware limits. We evaluated VibeSyn through three demonstrations of increasing scope and autonomy: evidence-to-action decision tasks, a ten-step total synthesis of catharanthine, and structure-to-compound workflows for cyclic peptides containing unnatural amino acids (uAAs). Across the demonstrated campaigns, active human input and simple human confirmation account for no more than 7.5% and 5.2%, respectively, of the recorded workflow actions and outcomes. VibeSyn therefore establishes a framework for end-to-end molecular synthesis in which synthetic intent, rather than predefined procedures, guides adaptive experimental decision-making.

## Vibe Synthesis system architecture

VibeSyn implements this intent-driven synthesis paradigm through a hierarchical architecture connecting LLM-based agents, computational tools, and laboratory automation (Fig. 2b). A synthesis campaign begins with a user-defined objective, which may specify a target structure or a synthesis procedure together with user-defined preferences, constraints, or knowledge. Rather than compiling these inputs into a fixed sequence of procedures, VibeSyn represents multistep synthesis as an evolving campaign state that is updated after each action. The system consequently advances through evidence-dependent transitions until the target compound is prepared and verified. A web application provides the interface for agent configuration, task submission, workflow monitoring, human intervention, and inspection of the synthesis history.

All LLM-based agents are instantiated through a unified configuration framework that specifies their task-specific role, knowledge, workflow, output schema, permitted recipients, and memory policy. The recipient set defines the collaboration topology of VibeSyn by routing each agent's output to designated agents, tools, devices, or human chemists. Agent memory is agent-specific and can be configured independently as either persistent or temporary mode: persistent memory retains interaction history and relevant experimental evidence across successive assignments, whereas temporary memory resets the context after an assignment is completed. This configurable policy preserves continuity when required, while limiting context growth and cross-task interference when persistent memory is unnecessary. Because agent behavior is defined through natural-language instructions, chemists can add, remove, or revise agent roles and decision rules through the web interface without rewriting task-specific orchestration code. The agent architecture is decoupled from the underlying LLM, enabling newly released, more capable models to be incorporated. This modularity allows advances in LLM capabilities to strengthen VibeSyn's chemical reasoning and experimental decision-making without redesigning the existing agent configuration framework. The task agents evaluated in this study primarily use OpenAI o3 (version o3-2025-04-16), selected during system development for its practical balance of cost and instruction-following performance.

Decision-making in VibeSyn is organized around a central MA and a set of function-specific task agents. The MA interprets the overall synthesis objective, assigns a chemically bounded task to an appropriate task agent, incorporates the returned result and associated evidence into the evolving campaign state, and determines the subsequent task from the updated synthesis status. Within each assignment, a task agent can invoke connected computational tools or automation devices through structured request-response interfaces, interpret the returned outputs, and determine whether further action is required before reporting its conclusion to the MA. VibeSyn herein comprises 10 task agents connected to 10 computational and multimodal analysis tools and 7 automated synthesis devices, spanning route and condition design, reaction development, downstream processing, product verification, and peptide synthesis (Fig. 2b-c). This hierarchical cycle is repeated throughout the synthesis campaign, allowing task-level reasoning and experimental evidence to be continuously integrated into campaign-level decisions. Human chemists participate through the same communication network when additional information, confirmation, or operations outside the configured chemical or hardware scope are required.

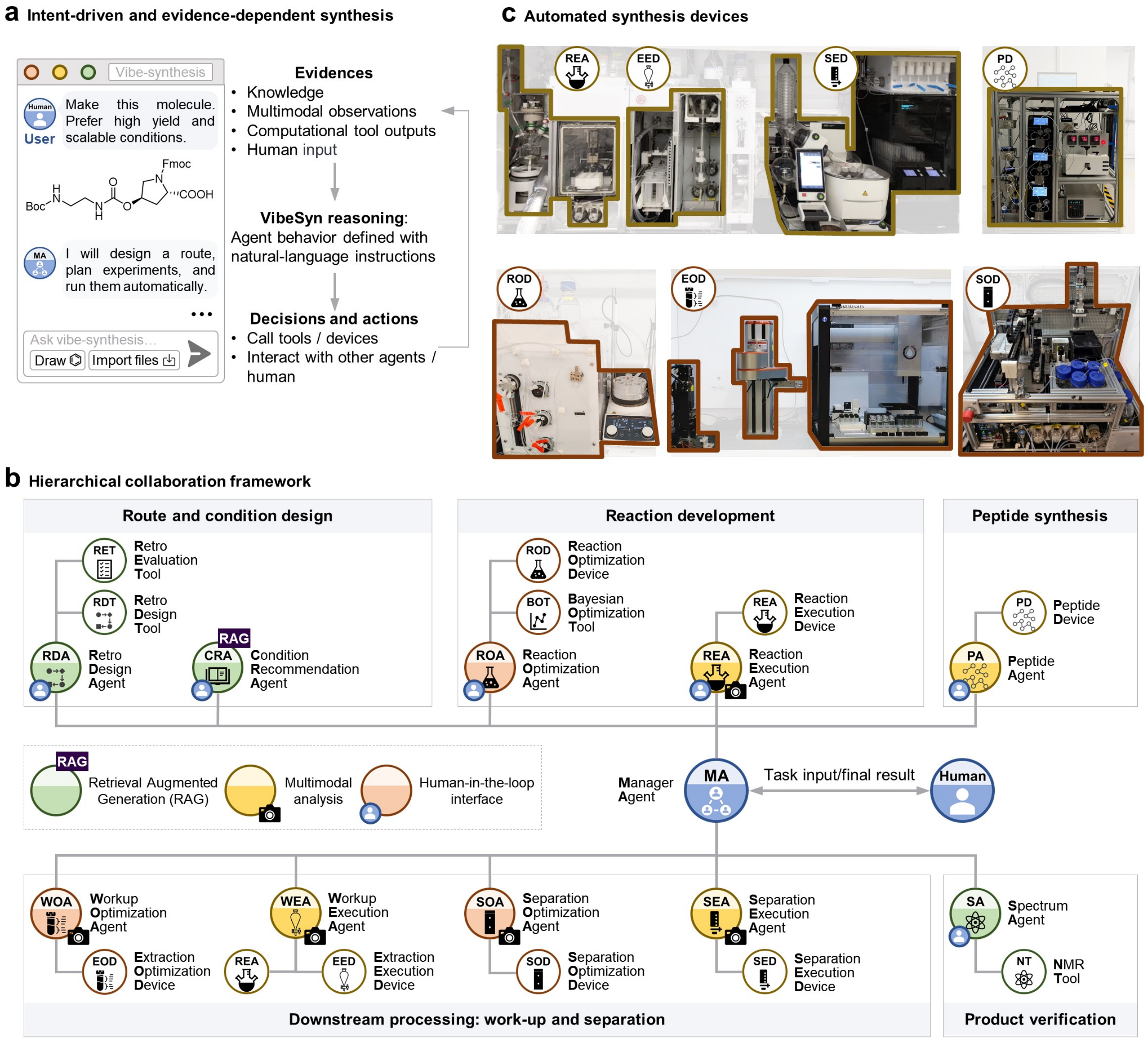

**Fig. 2 | VibeSyn architecture for end-to-end multistep synthesis. a**, Intent-driven and evidence-dependent synthesis workflow from user request to experimental execution and feedback. **b**, Hierarchical collaboration framework integrating task-specific agents, computational and multimodal tools, automated devices, and human chemists. **c**, Automated experimental devices for reaction, workup, separation, and peptide synthesis.

## Evidence-grounded synthesis decision-making

In long-horizon multistep synthesis, experienced chemists continuously integrate chemical knowledge with literature and computational evidence, interpret evolving physical and analytical observations, and weigh these inputs against practical constraints to determine the next course of action. VibeSyn formalizes this process as an evidence-to-action cycle in which task agents acquire and interpret relevant evidence, and translate that evidence into chemically actionable decisions (Fig. 2a). Next, we illustrate this evidence-to-decision process through three representative modes: knowledge-grounded reasoning, tool-augmented decision-making, and multimodal interpretation of experimental states.

Knowledge-driven decision-making

We used Fe-catalyzed ligand-to-metal charge-transfer (LMCT) C–H functionalization[37] to examine how VibeSyn combines parametric knowledge from the base LLM, quantitative knowledge retrieved from external references through retrieval-augmented generation (RAG), and procedural knowledge acquired from few-shot demonstrations (Fig. 3a). Drawing on its parametric knowledge, Condition Recommendation Agent (CRA) initially proposes the comparatively expensive decatungstate photocatalyst. To reduce catalyst cost and avoid ultraviolet irradiation, the user specifies the use of an inexpensive Fe catalyst. CRA consequently redirects the condition search toward $FeCl_3$- and $FeBr_3$-based LMCT systems operating under visible light.

Selecting between the two haloferrate systems requires a quantitative assessment of hydrogen-atom-transfer thermodynamics. The base LLM (version o3-2025-04-16) initially estimates the α-C–H bond-dissociation energy (BDE) of 1,4-dioxane (**2**) to be approximately 92 kcal $mol^{-1}$ based on its parametric knowledge of activated ether C–H bonds. Because this unsourced estimate directly affects the predicted reactivity of Cl and Br radicals, CRA retrieves quantitative data from established BDE handbooks via RAG[38,39]. The reference values place the α-C–H BDE of **2** at 96.0–96.6 kcal $mol^{-1}$, compared with approximately 103.1 kcal $mol^{-1}$ for H–Cl and 87.5 kcal $mol^{-1}$ for H–Br. CRA interprets that hydrogen abstraction from 1,4-dioxane is thermodynamically favorable for Cl radical but unfavorable for Br radical, and therefore prioritizes the chloride-based system.

After CRA returns the two candidate conditions to MA, MA assigns their experimental evaluation to Reaction Optimization Agent (ROA). Guided by few-shot demonstrations, ROA translates the corresponding natural-language procedures into structured JSON commands and dispatches them to Reaction Optimization Device (ROD), which executes the experiments and returns the measured yields. The $FeCl_3 \cdot 6H_2O/Bu_4NCl$ system affords the target product in 85.0% yield, whereas the $FeBr_3/Bu_4NBr$ system only gives 3.5% yield. These experimental outcomes are consistent with the thermochemical assessment and support selection of the $FeCl_3$-based condition.

Tool-augmented decision-making

Computational tools are invoked to augment task-agent reasoning when synthesis decisions require specialized capabilities that the base LLM alone cannot provide reliably (Fig. 3b). We demonstrated tool-augmented decision-making through the synthesis of an intermediate (**6**) for $Na_V1.7$ inhibitors[40]. After submitting the compound synthesis task to VibeSyn, MA calls Retrosynthesis Design Agent (RDA) to design the synthetic routes, and RDA invokes an in-house template-based Retrosynthesis Design Tool (RDT, ChemLex RetroPlanner). An LLM-based Retrosynthesis Evaluation Tool (RET) further compares and ranks the candidates based on safety, functional-group compatibility, precedent yield, step count, and reagent economy. RDA gives three top ranked routes and the Mitsunobu route is selected for experimental evaluation for its simplicity and high yield.

MA subsequently assigns ROA to explore the optimal synthesis condition. Initial screening of $PPh_3$/DIAD and P(*t*-Bu)$_3$/ADDP reagent combinations gives 17.5% and 23.1% yields, respectively, indicating the need for substantial optimization. The ROA then helps determine the optimization variables and then orchestrates the iterative optimization by invoking Bayesian Optimization Tool (BOT) and ROD. In each cycle, BOT uses accumulated condition–outcome data to propose new combinations of temperature, reagent identity, and reagent loading; ROD executes the proposed experiments and returns the measured yields; and ROA evaluates whether further exploration is warranted. Seven BOT-guided optimization rounds converge to the $PMePh_2$/ADDP condition that affords 81 ± 1% yield. Reaction Execution Agent (REA) subsequently executes the reaction at 4-mmol scale, followed by workup and purification by Workup Execution Agent (WEA) and Separation Execution Agent (SEA) to afford 0.95 g of the product in 76.6% isolated yield.

The purified product then requires structural verification. The Spectrum Agent (SA) invokes the NMR Tool (NT), which can process raw experimental data into structured NMR reports via Bruker's TopSpin API and predict NMR shifts using nmrshiftdb predictor[41]. The SA first compares processed experimental $^1$H NMR resonances with the literature spectrum and finds that the four aromatic proton groups differ by no more than 0.1 ppm. It assigns the overlapping $OCH_3$ and $NCH_2$ resonances to the combined five-proton signal and confirms that the isobutyl methine and methyl signals exhibit the expected multiplicities and integrations. The SA then compares the experimental $^{13}$C resonances with the predicted spectrum and identifies agreement for most signals. It identifies the brominated C3 carbon as the principal deviation, appearing at 89.8 ppm rather than the predicted value near 107 ppm. Rather than classifying this difference as a structural mismatch, the SA considers the known difficulty of predicting shifts for carbons directly bonded to heavy halogens. The SA therefore concludes that the complete $^1$H and $^{13}$C resonance patterns support the target structure.

Multimodal-interpretation decision-making

Many experimental decisions depend on physical states that are not captured by textual context alone, including reaction appearance, phase behavior, TLC patterns, and chromatographic traces. VibeSyn therefore integrates multimodal interpretation with task-specific agents: observable features are extracted from images and evaluated against chemistry-aware criteria (Fig. 3c). We demonstrated this capability using the $FeCl_3$-catalyzed photochemical alkylation of benzothiazole, in which multimodal observations guide reaction monitoring, workup, and purification.

Because the reaction follows a published procedure, MA directly assigns execution to REA. However, photochemical kinetics are sensitive to reactor geometry and irradiation conditions, making the literature reaction time difficult to transfer directly. In the reported Fe-LMCT mechanism[42], photoexcitation of Fe(III)-Cl generates Cl radicals and Fe(II), while Fe(II) is subsequently reoxidized to Fe(III) via single-electron transfer with the substrate-derived radical intermediate. Once conversion is complete, reoxidation terminates and less-colored Fe(II) accumulates. The REA therefore analyzes images of the reaction mixture every 2 h, and the observed transition from a yellow solution to an almost colorless suspension after 14 h is interpreted as a visual indicator of reaction completion.

For workup, Workup Optimization Agent (WOA) tests extraction of a 2-mL post-reaction aliquot with dichloromethane and water using the Extraction Optimization Device (EOD), producing an emulsion without a defined interface. The WOA interprets this state as consistent with colloidal $Fe(OH)_3$ and the high MeCN content, and therefore revises the conditions to a mixture of dichloromethane, brine, and 10% HCl. Brine promotes phase separation, whereas HCl dissolves residual $Fe(OH)_3$. The revised condition gives two clear phases with a well-defined interface, satisfying the phase-separation criterion. WEA then transfers the optimized protocol to Extraction Execution Device (EED) for full-batch extraction.

The Separation Optimization Agent (SOA) applies the same visual decision to TLC images from the Separation Optimization Device (SOD). With 10% ethyl acetate (EA) in petroleum ether, the product $R_f$ of 0.60–0.62 exceeds the predefined range of 0.30–0.50. Reducing the EA content to 5% gives acceptable $R_f$ of 0.42–0.46 for the product and 0.32–0.36 for residual benzothiazole, leading the SOA to specify a 5–10% gradient for flash chromatography. The resulting Separation Execution Device (SED) trace contains peaks at approximately 2.3 and 5.6 column volumes (CV). By combining these features with the gradient and TLC-derived polarity ordering, the SEA assigns the earlier peak to the product (**12**) and the later peak to residual substrate (**10**).

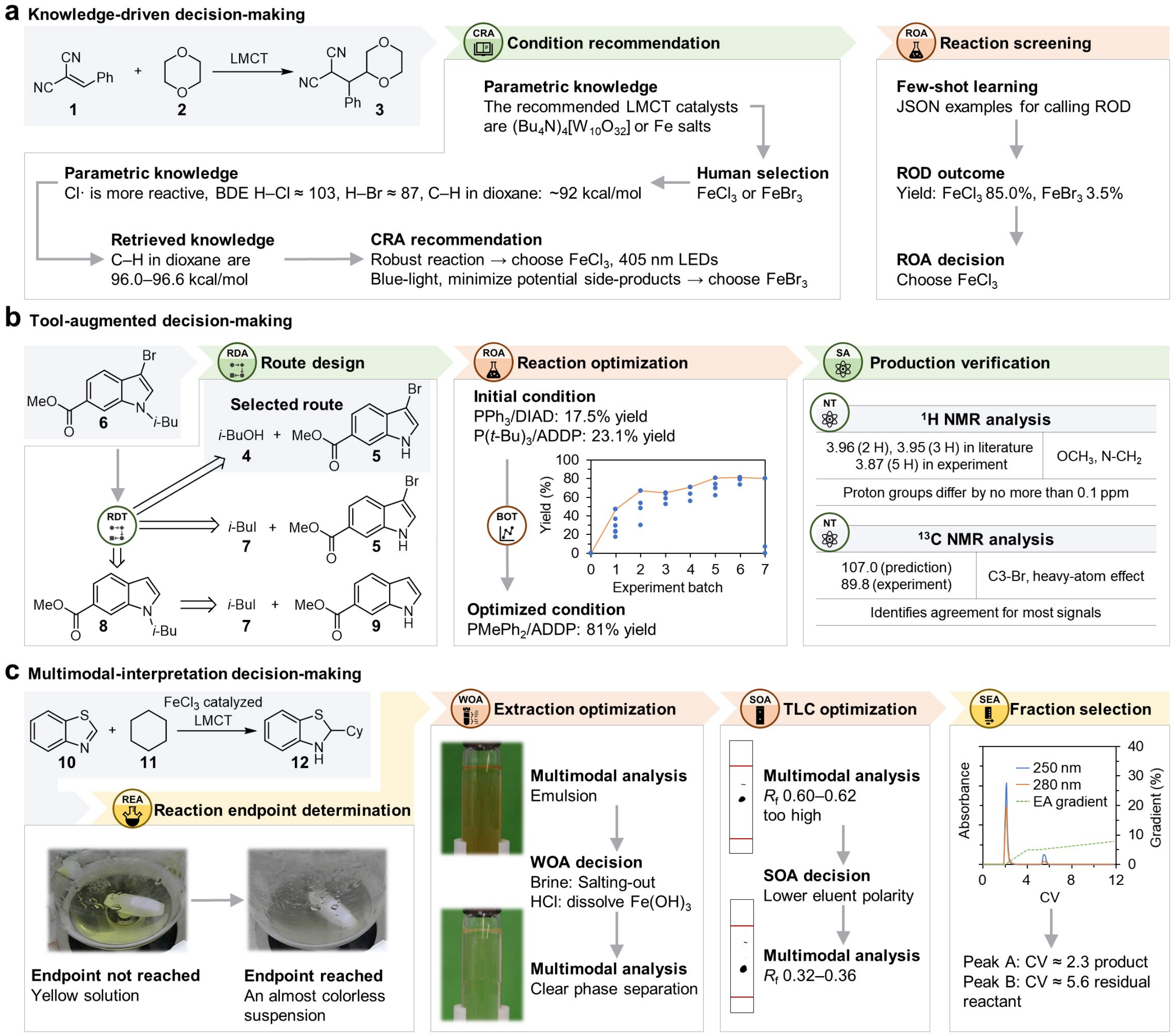

**Fig. 3 | Evidence integration and decision mechanisms in VibeSyn. a**, Knowledge-driven condition selection for Fe-catalyzed ligand-to-metal charge-transfer (LMCT) C–H functionalization. **b**, Tool-augmented route selection, Bayesian optimization, and NMR-based structure verification for Mitsunobu *N*-alkylation to synthesize the Nav1.7 inhibitor intermediate. **c**, Multimodal interpretation of reaction appearance, phase behavior, TLC, and flash-chromatography traces during $FeCl_3$-catalyzed photochemical alkylation. DIAD, diisopropyl azodicarboxylate; ADDP, 1,1′-(azodicarbonyl)dipiperidine.

## Decision-guided refinement of a ten-step total synthesis of catharanthine

Computational retrosynthetic planning can design experimentally validated routes to complex natural products, but executing these routes still requires operational, analytical, and strategic decisions that are rarely encoded in route-planning programs, leaving a gap for automated multistep synthesis of complex molecules[8]. To evaluate VibeSyn in this setting, we did not ask the system to design a new route, but instead tasked it with reproducing a ten-step synthesis of catharanthine (**24**), a key precursor to clinically important anticancer drugs, reported in 1987[43]. The original publication provided a chemically viable route but left substantial procedural detail implicit, and several transformations relied on older procedures for which more efficient modern alternatives are available. The campaign therefore provides a stringent test of whether VibeSyn could translate an established synthetic plan into an executable and adaptive laboratory workflow.

VibeSyn records 722 actions across the ten-step campaign, capturing decisions across a chemically dependent sequence from starting material to verified product. Starting from 29.8 mmol of 3-ethylpyridine (**13**), the campaign encompasses key transformations including Diels–Alder cycloaddition, photochemical cyclization, and chemoselective lactam reduction, furnishing 160 mg product (Fig. 4a). The records comprise 503 agent reasoning records (69.7%), 136 tool/device outcomes (18.8%), 53 active human inputs (7.3%), and 30 human confirmations (4.2%) (Fig. 4b). Active human inputs include task submission, selection or revision of agent-proposed conditions, apparatus-specific execution adjustments, and reporting of isolated product masses. Human confirmations occur at procedural, analytical, or safety checkpoints. VibeSyn thus handles most recorded actions, allowing chemists to focus on safety-critical, ambiguous, or strategic decisions. Representative agent decisions throughout the campaign are discussed below (Fig. 4c-d).

<u>Steps 1–2: Operational adaptation based on physical states</u>

The first two steps illustrate how agents convert physical observations into operational changes. In Step 1, REA analyzes the post-reaction image and identifies a milky suspension, and infers from the reaction context that the ionic quaternization product (**14**) is poorly soluble in acetone and could obstruct RED's transfer tubing and pump. REA therefore instructs RED to add ethanol, producing a clear, homogeneous solution suitable for automated transfer. In Step 2, WOA observes a persistent emulsion with an unclear interface during an EOD aliquot test of the reduction workup. WOA attributes this behavior to the ethanol, suspended inorganic salts, and the absence of a salting-out additive, and replaces the initial water extraction with a dichloromethane/brine system containing dilute HCl. The revised conditions promote phase separation and dissolve residual inorganic material, producing a well-defined phase boundary.

<u>Steps 3–4: Analytical interpretation for fraction selection and product verification</u>

Steps 3–4 demonstrate how analytical evidence guides purification and structural assessment. In Step 3, given the target's (**16**) carbamate and alkene groups and lack of an aromatic chromophore, SEA expects it to be more polar and substantially less UV-active than residual benzyl chloride, toluene, and N-benzylated substrate (**15**). SEA assigns the intense response at approximately 0.5–1.5 CV under 0% EA to the solvent front and residual aromatic components. At increased solvent polarity, it assigns the weak, broad response at approximately 5–7.5 CV to the target and directs SED to collect these fractions, affording 2.79 g of product in 91.8% isolated yield.

In Step 4, SA uses NT for spectral processing and $^{13}C$ NMR shift prediction, and receives literature $^{1}H$ NMR data from MA. The processed $^{1}H$ spectrum reproduces literature shift ranges, integrations, and multiplicities, including the diagnostic ethyl methyl, methoxy, and brominated methine resonances at 1.14, 3.74, and 4.61 ppm, respectively. SA attributes additional $^{13}C$ signals beyond the prediction to stereochemical nonequivalence in the dibrominated product (**17**) and notes the largest shift deviations at bromine-substituted carbons. SA concludes that the combined spectral patterns support the proposed structure.

<u>Steps 5–6: Reaction optimization towards safer conditions</u>

Steps 5 and 6 test whether experimental outcomes can redirect condition selection. To avoid carcinogenic HMPA solvent in the literature condition, ROA directs ROD to investigate alternative DBU in NMP and DMSO. Both conditions give approximately 97% conversion but only 8–9% target yield (**18**), leading ROA to identify selectivity as the limiting factor. ROA therefore proposes alternative bases and solvents, including DBU and DABCO in MeCN. ROD returns target yields of 27.6% and 47.7%, respectively. Although DABCO only gives 55.9% conversion, the target is the major product. ROA interprets this result as improved selectivity and increases the DABCO loading. A subsequent ROD experiment using five equivalents of DABCO reaches complete conversion and a 90.3% yield. REA transfers the optimized condition to RED for scale-up. The literature procedure supplied to VibeSyn explicitly reports that the resulting 1,6-dihydropyridine is unstable, and MA therefore incorporates this constraint into the campaign decision to omit purification and direct the crude **18** for the following Diels–Alder cycloaddition in Step 6. The two-step sequence affords 1.75 g of cycloadduct (**19**) in 44.1% isolated yield.

<u>Step 7: Safety-bounded human participation</u>

Step 7 demonstrates how VibeSyn handles a transformation outside the configured automation envelope. CRA identifies in situ generation of trimethylsilyl iodide (TMSI) from iodine and hexamethyldisilane, requiring handling of corrosive and moisture-sensitive iodine reagents beyond the available automated hardware's capabilities. CRA therefore recommends manual execution, and MA assigns the transformation to a human chemist. During this intervention, VibeSyn retains responsibility for condition planning, downstream coordination, and subsequent evidence-based decisions.

Steps 8–9: Downstream-aware purification

Steps 8 and 9 show how purification decisions are evaluated against downstream requirements rather than local purity alone. During step 8, SOA analyzes TLC images generated on SOD, evaluates the target $R_f$ and tailing of **22**, and iteratively adjusts the eluent. Increasing the EA content from 40% to 50% moves the target into the desired $R_f$ range, whereas adding 1% triethylamine at 55% EA reduces but does not eliminate tailing and overlap. After three attempts, SOA reports to MA acceptable target retention but incomplete resolution from a minor, less-polar impurity. Because TLC indicates that the target is the predominant component, MA judges that further optimization would risk material loss without sufficient improvement in purity. MA therefore authorizes a rough chromatographic cut and defers precise separation after Step 9. Following the $Ir(ppy)_3$-catalyzed photochemical cyclization under 450-nm irradiation, SOA identifies petroleum ether/EA/triethylamine (60:39:1) as a suitable eluent. SEA directs SED to execute the flash separation, affording 210 mg of 5-oxo-catharanthine (**23**).

Step 10: Chemoselective redesign of the lactam reduction

The final step tests route-level redesign rather than local optimization. The 1987 literature protocol converts **23** into **24** through a multistep thionation, *S*-alkylation, and reduction sequence involving Lawesson's reagent, methyl iodide, and sodium cyanoborohydride. VibeSyn instead evaluates whether the lactam can be reduced directly while retaining the ester. CRA compares candidate one-step reduction strategies by considering their chemoselectivity and operational trade-offs. The human chemist selects Rh-catalyzed hydrosilylation on the basis of literature precedent. CRA specifies $RhH(CO)(PPh_3)_3$ and phenylsilane in THF, and REA converts the selected condition into a scale-specific protocol and directs RED to execute the reduction. Reduction of 0.50 mmol of **23** affords 160 mg of **24** in 95% isolated yield, with NMR supporting product identity and ester retention.

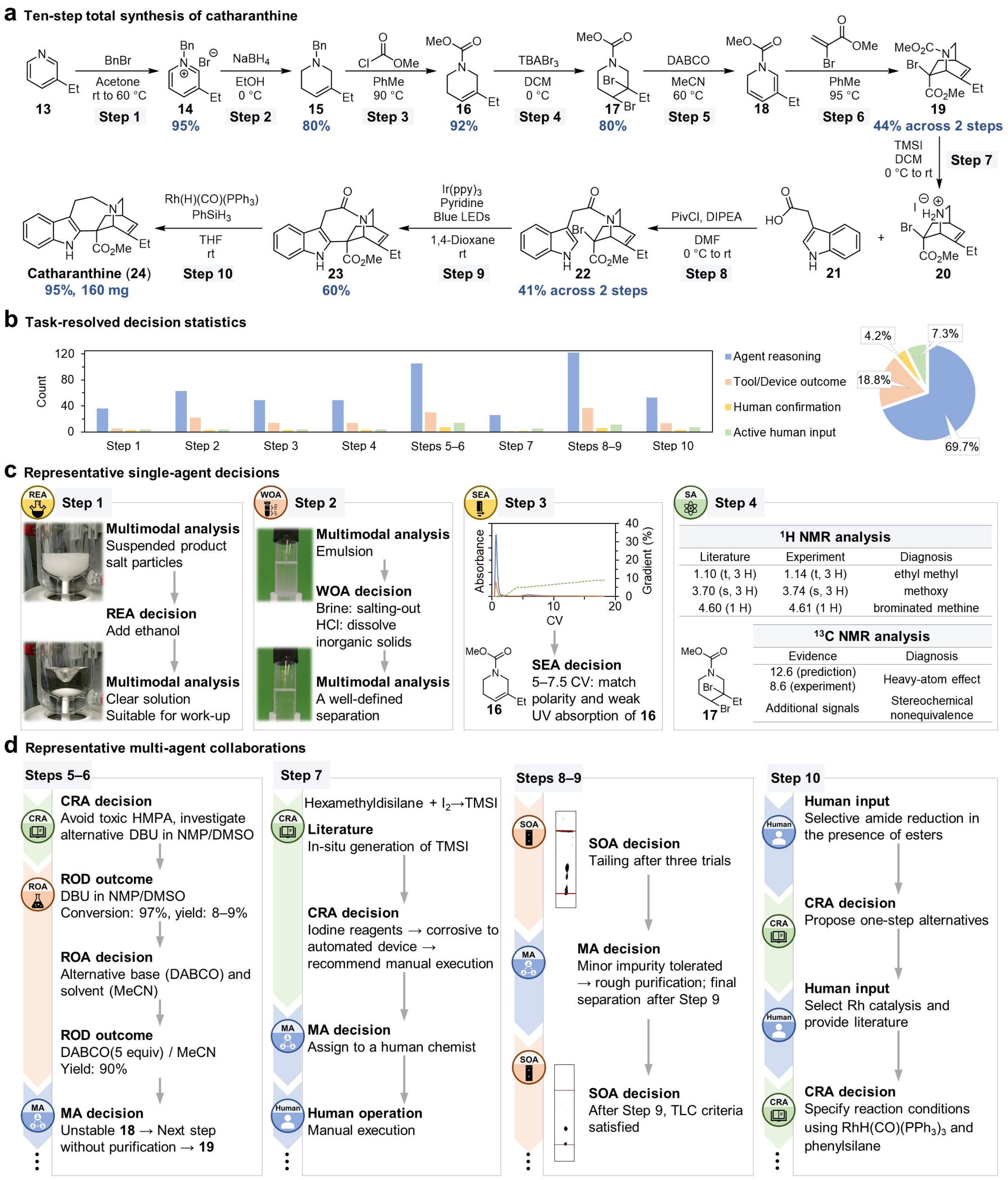

**Fig. 4 | Decision-guided refinement of a ten-step total synthesis of catharanthine. a**, Ten-step synthetic route refined by VibeSyn system for preparation of 160 mg catharanthine. **b**, Step-resolved counts and overall fractions of agent reasoning, tool or device outcomes, human confirmation, and active human input. **c**, Representative single-agent decisions for operational adaptation, analytical interpretation. **d**, Representative multi-agent collaborations for reaction optimization, safety-bounded human participation, downstream-aware purification, and chemoselective reduction throughout the ten-step synthesis campaign. BnBr, benzyl bromide; TBABr$_3$, tetrabutylammonium tribromide; DABCO, 1,4-diazabicyclo[2.2.2]octane; PivCl, pivaloyl chloride; HMPA, hexamethylphosphoramide.

## Structure-to-compound end-to-end synthesis of cyclic peptides

At the third and broadest level of evaluation, we examined VibeSyn in structure-to-compound synthesis, where a chemist specifies the target and VibeSyn determines and executes the automated workflow required to prepare and verify it. Cyclic peptides incorporating uAAs have attracted growing therapeutic interest because uAAs substantially expand accessible chemical and conformational space, yet their development is often constrained by non-commercial monomer preparation and integration into peptide-synthesis workflows. We therefore selected uAA-containing cyclic peptides to evaluate VibeSyn's structure-to-compound synthesis capability. Starting from a peptide sequence and uAA structures as SMILES strings, VibeSyn was tasked with designing and synthesizing uAAs at sufficient scale, then integrating them into DAN linker-based solid-phase peptide synthesis (SPPS)[44,45] and macrocyclization workflows (Fig. 5a). The three campaigns progressively extend decision-making from monomer-scale planning and peptide-coupling refinement to route-level traceback and symmetry-aware library construction (Fig. 5c–e). Across these campaigns, agent reasoning (72.5%) and tool/device outcomes (14.7%) together account for 87.2% of the recorded actions and outcomes; active human input and human confirmation account for 7.5% and 5.2%, respectively (Fig. 5b).

### uAA synthesis and peptide-coupling refinement

The first campaign targets pasireotide (**29**), a uAA-containing cyclic hexapeptide and multireceptor somatostatin analogue used clinically for Cushing disease and acromegaly[46,47]. Its compact macrocyclic architecture and epimerization-prone phenylglycine (Phg) residue enable evaluation of how VibeSyn coordinates uAA synthesis with cyclic-peptide assembly. The chemist provides MA with the sequence Phe–$X_1$–Phg–D-Trp–Lys–Tyr(Bn) and the SMILES of Fmoc-$X_1$-OH (Fmoc-L-Hyp(Bom)-OH, **28**). MA initiates uAA preparation, and RDA selects a concise two-step route that temporarily masks the carboxylic acid as a methyl ester and limits functional-group competition during side-chain functionalization. MA sets the starting scale at 5.6 mmol of **25**, and the two-step synthesis affords 1.58 g of **28** in 52.3% overall yield.

For cyclic peptide synthesis using the resulting uAA, the Peptide Agent (PA) draws on peptide-design knowledge, which prioritizes *N*-terminal nucleophilicity, resin-proximal *C*-terminal steric hindrance, and protecting-group compatibility. PA evaluates the six possible ring disconnections and selects the Tyr(Bn)–Lys(Boc) junction, giving the linear sequence Tyr(Bn)–Phe–$X_1$–Phg–D-Trp(Boc)–Lys(Boc). This junction combines an accessible *N*-terminal primary amine with moderate *C*-terminal steric demand while avoiding $X_1$'s secondary amine and sterically hindered Phg at unfavorable termini. Initial synthesis using Peptide Device (PD) produces two resolved HPLC peaks with approximately equal UV areas and identical ESI-MS ions matching the calculated mass. PA combines these observations with the absence of truncation or adduct ion series and Phg's susceptibility to base-promoted epimerization under HATU/DIPEA, identifying the Phg-containing coupling as the most likely origin of the two peaks. To reduce epimerization, PA recommends using COMU/2,6-lutidine and a milder temperature at the Phg–D-Trp(Boc) junction while retaining HATU/DIPEA for the remaining couplings. The revised conditions produce a dominant HPLC peak (>95% UV area).

Traceback-guided route redesign for a bicyclic peptide

Octreotide, a clinically established, long-acting cyclic somatostatin analogue, provides a pharmacologically validated scaffold for structural diversification[48]. Incorporating a $CF_3$-pyrazole-bearing uAA and a second cyclization provides an evaluation of VibeSyn's ability to coordinate heterocyclic monomer synthesis with topology-controlled bicyclic-peptide assembly[49]. We tasked MA with the target sequence D-Phe–Cys–Phe–D-Trp–$X_2$–Thr–Cys–Thr, and the structure of Fmoc-protected monomer X (**40**). RDA generates and evaluates ten candidate routes and prioritizes a four-step sequence beginning with a Knorr condensation between a $CF_3$-substituted 1,3-dicarbonyl precursor (**30**) and an arylhydrazine (**31**). In three trials varying temperature and solvent, ROA obtains 0% of the desired product (**36**) and consistently forms the undesired 3-$CF_3$/5-phenyl regioisomer (**32**). ROA attributes this to preferential terminal $NH_2$ condensation at the more electrophilic $CF_3$-adjacent carbonyl. ROA reports the failure to MA, which traces it to route design and reassigns the task to RDA. RDA evaluates the supplied literature precedents and develops a milder six-step β-methoxy-enone route. Its pivotal [3+2] cyclocondensation between the β-methoxy $CF_3$ enone (**34**) and an arylhydrazine (**31**) produces a dihydropyrazol-5-ol intermediate (**35**), which is subsequently dehydrated to the desired pyrazole regioisomer (**36**). MA coordinates the redesigned sequence, affording 1.89 g of crude Fmoc-protected monomer (**40**) in 15.4% overall six-step yield.

MA then assigns the peptide-design task to PA. PA evaluates all possible cyclization junctions and recommends the D-Phe–Thr junction for its accessible *N*-terminal primary amine and limited *C*-terminal steric hindrance near the resin. PA submits the linear sequence to PD for assembly and head-to-tail cyclization on DAN resin. LC-MS analysis confirms the formation of the expected monocyclic peptide. To establish the second ring, PA recommends $I_2$-mediated oxidation to form the intramolecular Cys–Cys disulfide bond. The corresponding −2 Da mass shift confirms the bicyclic product, which is obtained at greater than 95% purity after preparative flash chromatography.

Parallel construction of a peptide library

Gramicidin S is a $C_2$-symmetric cyclic decapeptide antibiotic comprising two Val–Orn–Leu–D-Phe–

Pro repeats, and its compact β-sheet conformation makes it a well-established scaffold for residue-level diversification[50,51]. Its repeated architecture and two symmetry-related Orn positions provide a benchmark for coordinating monomer preparation, symmetry-aware library design, and cyclic-peptide synthesis. MA receives the analogue scaffold Val–$X_3$–Leu–D-Phe–Pro–Val–$X_4$–Leu–D-Phe–Pro and the structures of five ε-acylated Lys building blocks (Fig. 5e). MA assigns RDA to design syntheses of the three non-commercial uAA monomers. RDA proposes direct ε-acylation of Fmoc-Lys-OH, using acid-chloride acylation for the ε-benzoyl and ε-(4-fluorobenzoyl) derivatives, and HATU-mediated coupling for the ε-nicotinoyl derivative. MA coordinates execution of these routes, affording Fmoc-Lys(Bz)-OH (**47**), Fmoc-Lys(4-FBz)-OH (**48**), and Fmoc-Lys(Nic)-OH (**49**) in 68%, 65%, and 57% yield, respectively. MA adds commercial Fmoc-Lys(Tfa)-OH (**50**) and Fmoc-Lys(Ac)-OH (**51**) and transfers all five monomers to PA.

PA recognizes that $X_3$ and $X_4$ are interchangeable under the scaffold's twofold rotational symmetry. It therefore reduces the nominal 5 × 5 ordered matrix from 25 combinations to 15 nonredundant cyclic peptides: five homosubstituted and ten heterosubstituted analogues. PA evaluates the candidate cyclization junctions and implements the D-Phe–Leu junction across the library. This arrangement places an accessible D-Phe primary amine at the *N* terminus and Leu at the *C* terminus while avoiding Pro and the variable Lys derivatives at the cyclization termini. PA subsequently generates and dispatches the 15 SPPS and head-to-tail cyclization workflows to PD and evaluates each product by LC-MS. All 15 cyclic peptides meet PA's quality criteria, with dominant chromatographic peaks, >95% HPLC purity, and expected masses.

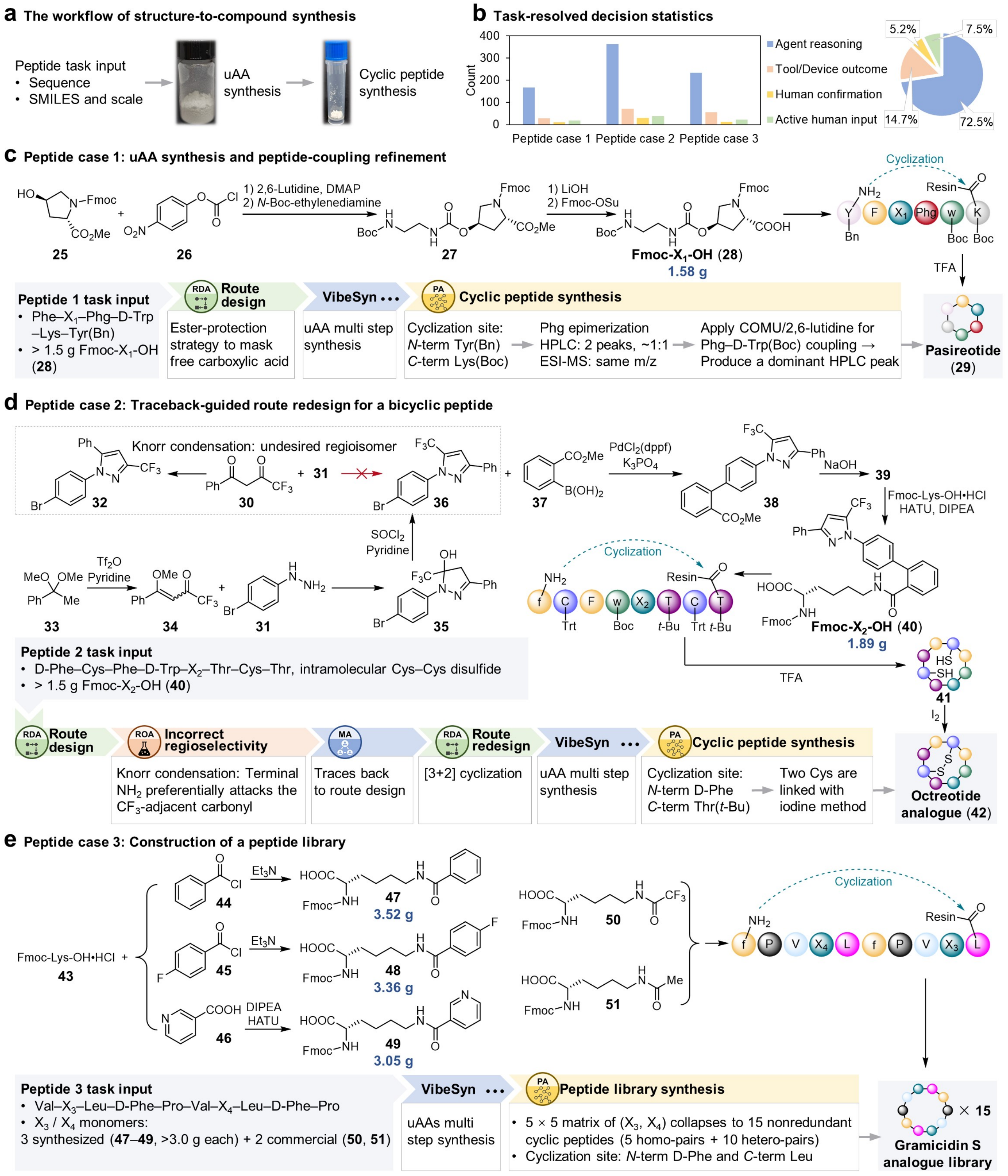

**Fig. 5 | Structure-to-compound synthesis of cyclic peptides containing unnatural amino acids. a**, Workflow from peptide-task specification to monomer preparation and cyclic peptide synthesis. **b**, Task-resolved counts and overall fractions of agent reasoning, tool or device outcomes, human confirmation, and active human input. **c**, Monomer synthesis, macrocyclization-site selection, and coupling-condition refinement for solid-phase assembly and head-to-tail cyclization of the cyclic hexapeptide pasireotide. **d**, Traceback-guided route redesign for $CF_3$-pyrazole monomer synthesis and assembly of an octreotide-derived bicyclic peptide by head-to-tail cyclization and disulfide formation. **e**, Parallel synthesis of three substituted lysine monomers and symmetry-aware construction of a 15-

member library of gramicidin S cyclic-decapeptide analogues through SPPS and head-to-tail cyclization. DMAP, 4-dimethylaminopyridine; COMU, (1-cyano-2-ethoxy-2-oxoethylidenaminooxy)dimethylaminomorpholinocarbenium hexafluorophosphate. Uppercase and lowercase single-letter amino acid codes denote L- and D-amino acids, respectively (e.g., F = L-Phe and f = D-Phe).

## Discussion

VibeSyn provides a hierarchical architecture for adaptive multistep synthesis that translates a chemist's target-level intent into evidence-dependent experimental actions. MA maintains the campaign objective and evolving experimental state, while task agents integrate chemical knowledge, computational tools, multimodal observations, and automated devices to make bounded decisions and direct experimental operations. This organization enables VibeSyn to respond to new evidence by continuing execution, correcting operational failures, optimizing tactical choices, or revising upstream routes. Demonstrations include a ten-step total synthesis of catharanthine and structure-to-compound cyclic-peptide campaigns. Across these campaigns, active human input and simple confirmation account for at most 7.5% and 5.2% of the recorded actions, respectively; agent reasoning and tool/device outcomes account for more than 87%. The central advance is enabling accumulated experimental evidence to adaptively redirect the synthesis workflow.

Current capabilities remain bounded by configured task-agent expertise and the operating ranges of connected tools and devices. Decision reliability depends on evidence quality and how faithfully the campaign state represents the experiment. Future task agents should acquire stronger domain expertise by coupling updatable chemical knowledge to validated tools and an explicit understanding of what each device can safely and reliably perform. They should explicitly represent uncertainty so ambiguous evidence triggers additional measurements or targeted human input. Experimental-state representation should likewise progress from isolated images and analytical traces toward continuous multimodal evidence from time-resolved video, in situ spectroscopy, and equipment data. Such information would give task agents a more complete view of chemical and apparatus changes.

Greater decision autonomy must be matched by greater physical flexibility. Reconfigurable robotics capable of manipulating diverse materials and adapting apparatus under controlled conditions would expand VibeSyn beyond its current task-specific devices' capabilities. These advances should increase the fraction of actions completed without human input while retaining explicit authority and safety boundaries. Industrial process development will further require the campaign state to capture scale-dependent changes in reaction performance, process safety, and equipment behavior. Integrating process analytical technologies and pilot-scale batch or continuous equipment could extend VibeSyn from target synthesis to developing robust, safe, and scalable production processes.

## Methods

### Vibe Synthesis system architecture

VibeSyn provides a hybrid collaboration environment in which LLM-based agents, computational and multimodal tools, automated devices, and human chemists communicate through a shared task interface. A LangChain-based layer coordinates agent calls and routes messages among permitted recipients, whereas standardized adapters connect the task agents to computational tools and device-control services. A web application supports agent, tool, and device configuration, task submission, campaign monitoring, human intervention, and inspection of the synthesis history.

A synthesis campaign encompasses the complete process from submission of a synthesis objective to preparation and verification of the target compound or termination of the workflow. A task is a chemically bounded objective assigned by MA to a task agent. A workflow record is a completed message, action, or outcome generated by an agent, computational tool, automated device, or human chemist. Each record is stored chronologically in the campaign state together with the current task, accumulated evidence, pending operations, and links to generated resources.

The communication network, message formats, permitted recipients, and operating boundaries of tools and devices are configured in advance. Within these boundaries, the sequence of experimental procedures can evolve during a campaign: MA and the task agents use the accumulated evidence to determine whether to advance the synthesis, revise or repeat the current step, return to an upstream task, or request human input. The campaign terminates when the target satisfies the predefined preparation and verification criteria or when no further executable task remains within the configured scope.

Agent configuration and coordination

Each LLM-based agent is defined by a system prompt specifying its role, task-specific chemical knowledge, workflow instructions, and output format. Its configuration also specifies the base LLM, memory policy, and permitted recipients, including other agents, computational tools, automated devices, and human chemists. Requests to tools and devices are formatted as structured JSON commands, whereas agent and human communications follow recipient-specific response formats. Agent memory can be persistent, retaining relevant history across successive tasks, or temporary, resetting the active context after task completion while preserving the exchange in the campaign history. The task agents used in the reported campaigns primarily employ OpenAI o3 (model snapshot o3-2025-04-16). Each agent is specified as follows.

**Manager Agent (MA).** MA maintains the campaign-level objective and accumulated evidence. It decomposes the synthesis campaign into chemically bounded tasks, assigns each task to the appropriate task agent, and incorporates the returned result into the campaign state. MA then determines whether the next task should advance the synthesis, revise the current step, or return to an upstream decision.

**Retrosynthesis Design Agent (RDA).** RDA receives a route-design task from MA and submits the target SMILES to the Retrosynthesis Design Tool (RDT). It invokes the Retrosynthesis Evaluation Tool (RET) to compare the generated routes and interprets the returned evidence to select a route consistent with the synthesis objective and constraints. When MA returns experimental evidence indicating an intrinsic route limitation, RDA incorporates that evidence into the design of an alternative route.

**Condition Recommendation Agent (CRA).** CRA converts a selected route or literature procedure into an experimentally actionable specification. It evaluates the transformation against the available hardware, incorporates user-defined constraints, standardizes the reaction scale and procedure, and determines whether reaction, workup, or separation optimization is required. When the decision requires quantitative external knowledge, CRA uses retrieval-augmented generation (RAG) to retrieve and interpret relevant reference data.

**Reaction Optimization Agent (ROA).** ROA translates a candidate procedure into small-scale experiments and determines whether the resulting performance satisfies the synthesis objective. For multivariable optimization, ROA defines the design space and invokes the Bayesian Optimization Tool (BOT) to propose new conditions. It converts the proposed conditions into executable commands for the Reaction Optimization Device (ROD), integrates the returned yields into the condition–outcome history, and determines whether to continue optimization, select a condition, or report a reaction-level failure to MA.

**Reaction Execution Agent (REA).** REA scales the selected reaction procedure, calculates the required material quantities, and generates structured commands for the Reaction Execution Device (RED). During execution, REA can invoke a multimodal image-analysis tool to evaluate the observable state of the reaction mixture. It interprets the returned features against task-specific criteria to determine whether to continue the reaction, initiate quenching, or modify the post-reaction handling.

**Workup Optimization Agent (WOA).** WOA evaluates candidate workup conditions using an aliquot of the post-reaction mixture on the Extraction Optimization Device (EOD). It interprets multimodal evidence of phase separation or emulsification together with the composition of the mixture and revises the extraction conditions when necessary. Once the phase-separation criterion is satisfied, WOA returns the optimized solvent and additive ratios for full-batch processing.

**Workup Execution Agent (WEA).** WEA translates the selected workup procedure into scaled transfer and extraction operations. It directs the movement of the post-reaction mixture through tubing and pumps from RED to the Extraction Execution Device (EED) and controls solvent addition, mixing, settling, phase transfer, drying, and equipment cleaning. Images and conductivity measurements returned by EED allow WEA to determine whether phase separation and transfer are complete.

**Separation Optimization Agent (SOA).** SOA converts an initial separation proposal into TLC experiments on the TLC Optimization Device (SOD). It interprets the spot positions, shapes, and estimated retention factors returned by the multimodal tool against predefined retention and resolution criteria. SOA adjusts the mobile-phase composition until these criteria are satisfied and returns the optimized flash-chromatography method to MA and SEA.

**Separation Execution Agent (SEA).** SEA receives the optimized separation method from SOA and converts it into rotary-evaporation and flash-chromatography commands for the Separation Execution Device (SED). It interprets the chromatographic trace together with the programmed gradient, expected crude composition, and TLC-derived polarity ordering to identify fractions provisionally assigned to the target. Definitive product identity is established through subsequent analytical verification.

**Spectrum Agent (SA).** SA evaluates whether the isolated material is consistent with the proposed structure. It first invokes the NMR Tool (NT) to convert the raw experimental spectra into a structured report and, when required, generate predicted chemical shifts. SA then compares the processed experimental signals with literature or predicted reference spectra, evaluates chemically interpretable deviations, and returns its product-identity assessment to MA.

**Peptide Agent (PA).** PA receives the target peptide sequence and the structures of any required uAAs. It evaluates protecting-group compatibility, possible ring disconnections, coupling requirements, and target topology to select a linear precursor, cyclization junction, and coupling strategy. PA converts the resulting design into commands for the Peptide Device (PD) and interprets the returned HPLC-MS data to determine whether peptide assembly or cyclization requires revision.

<u>Human oversight and participation</u>

Human chemists participate as permitted recipients within the VibeSyn collaboration network. An agent can request human input when additional information or operational guidance is required, when a proposed task exceeds the configured chemical or hardware scope, or when confirmation is required at a procedural, analytical, or safety-related checkpoint. Human chemists can also inspect the campaign history, provide additional information, revise a pending instruction, stop the campaign, or return it to an earlier recorded state through the web application.

Active human input refers to a human-authored message that introduces substantive information or alters the synthesis workflow, including task submission, selection or revision of an agent-proposed condition, apparatus-specific adjustment, or reporting of an experimental outcome unavailable from a connected device. Simple human confirmation refers to a brief approval of a proposed procedure, analytical interpretation, or safety-related action that introduces no additional scientific or operational information.

<u>Computational and multimodal tools</u>

Computational and multimodal tools are registered in VibeSyn through a unified configuration interface that specifies each tool's function, permitted calling agents, input schema, and output schema. A task agent invokes a tool through a structured request, and the returned result is recorded as evidence in the campaign state before the agent determines the subsequent action. Each tool is specified as follows.

**Retrosynthesis Design Tool (RDT).** RDT uses the in-house template-based ChemLex RetroPlanner to generate candidate synthesis routes from a target SMILES. The search applies reaction templates derived from patent procedures and returns routes containing the proposed reactant structures, reaction sequences, and associated experimental precedents.

**Retrosynthesis Evaluation Tool (RET).** RET uses OpenAI o3 (model snapshot o3-2025-04-16) to compare routes supplied in structured JSON format. Its task-specific prompt evaluates safety and practical feasibility, functional-group compatibility, retrosynthetic logic, precedent yield, step count, and reagent economy. RET returns a comparative assessment to RDA.

**Retrieval-augmented generation (RAG).** RAG provides quantitative chemical information from a defined collection of reference handbooks. The documents are parsed using PyMuPDF and LlamaParse, segmented using the LlamaIndex SentenceSplitter, embedded using text-embedding-3-small, and indexed using VectorStoreIndex. For each query, RAG returns relevant passages and their source information to the requesting agent.

**Bayesian Optimization Tool (BOT).** BOT proposes reaction conditions from the design space specified by ROA and the accumulated condition–outcome data. It uses a Gaussian-process surrogate with a mixed kernel comprising a Matérn-5/2 kernel for continuous variables and a categorical kernel for discrete variables. Expected Improvement is used as the acquisition function, and multiple experimental conditions can be proposed within a single optimization round.

**NMR Tool (NT).** NT integrates the Bruker TopSpin Python interface and the nmrshiftdb prediction service. TopSpin routines convert the experimental peak and integration data into a solvent-filtered and normalized report containing chemical shifts, integrations, multiplicities, coupling constants, and consolidated $^{13}C$ signals. The proposed molecular structure is separately submitted to the nmrshiftdb API as a SMILES string, which returns predicted $^{1}H$ and $^{13}C$ chemical shifts for comparison with the experimental spectrum. NT returns both the processed experimental report and predicted values to SA.

**Multimodal image-analysis tools.** Five task-specific multimodal tools use GPT-5 (model snapshot gpt-5-2025-08-07) to process images of reaction mixtures, extraction states, TLC plates, and flash-chromatography traces. Each tool receives an image or rendered trace together with a task-specific prompt defining the relevant observable features and evaluation criteria. The structured output is returned to the corresponding task agent, which interprets it together with the chemical context to determine the next action.

Automated devices and control software

Automated devices are connected to VibeSyn through device-specific control services. High-level control of ROD, RED, EOD, EED, SOD, and SED is implemented in LabVIEW over a centralized TCP/IP network, whereas PD uses its previously reported control software. Agents communicate with the devices through RESTful HTTP interfaces using structured JSON requests and responses. A scheduler parses each request and distributes its operations to device-specific workers, which communicate with pumps, valves, motors, sensors, and commercial instruments through TCP/IP or Ethernet-to-serial interfaces supporting RS485, RS232, and CAN protocols. Device operations run asynchronously, and their status and experimental outputs are returned to the requesting agent and stored in the campaign state. Device-monitoring data are recorded in a MongoDB database. Each device is specified as follows.

**Reaction Optimization Device (ROD).** ROD comprises reaction vials mounted on temperature-controlled stirring positions, a syringe pump, selector valves, three-way solenoid valves, and reservoirs containing solvents or reagent stock solutions. A shared gas and liquid network connects the reservoirs and reaction vials to nitrogen and vacuum lines. The solenoid and selector valves establish the required atmosphere and fluidic pathway, after which the syringe pump sequentially dispenses the specified stock solutions into the selected reaction vial. The fluidic lines are cleaned automatically between experiments to limit cross-contamination.

**Reaction Execution Device (RED).** RED is built around a multiport jacketed glass reactor connected to a circulating temperature controller, nitrogen and vacuum lines, a condenser, automated liquid- and solid-dosing modules, a camera, and a material-discharge system. An independent magnetic drive controls stirring, whereas a torque motor actuates the threaded plug that opens and closes the reactor discharge port. Liquid reagents are stored as stock solutions in sealed vials within a nitrogen-protected enclosure and delivered by a syringe-pump-driven three-axis handler. Solid reagents are enclosed in tetramethylbenzene capsules and introduced through a 12-position dosing module. The camera acquires images of the reaction mixture at agent-specified intervals for multimodal analysis.

**Extraction Optimization Device (EOD).** EOD combines a SCARA robotic arm, an Opentrons OT-2 liquid handler, an automated cap-opening module, and a fixed camera for small-scale extraction tests. The robotic arm transfers glass vessels among the liquid-handling, mixing, settling, and imaging positions. The cap-opening module provides access for solvent and additive delivery, after which the vessel is mixed, allowed to settle, and positioned in front of the camera. The acquired image is returned to WOA through the corresponding multimodal tool.

**Extraction Execution Device (EED).** EED uses a mechanically stirred extraction vessel connected to solvent reservoirs, a multichannel selector valve, peristaltic pumps, filtration and drying columns, a camera, and an inline conductivity sensor. Post-reaction mixtures and reactor rinses pass through a diatomaceous-earth filtration column before entering the extraction vessel. After solvent addition, mixing, and settling, the camera records the phase state. During transfer, the conductivity sensor distinguishes organic and aqueous liquids, allowing discharge to stop at the phase boundary specified by WEA. The collected organic stream then passes through a molecular-sieve drying column before transfer to SED.

**Separation Optimization Device (SOD).** SOD comprises TLC-plate and sample-storage platforms, solvent reservoirs, syringe and peristaltic pumps, a piezoelectric spotting valve, a developing chamber, an imaging station, and a two-axis plate-handling stage. The pumps prepare the specified mobile phase and deliver the sample to the spotting valve. The two-axis stage transfers the plate sequentially through sample application, development, and imaging. The resulting image is returned to SOA for estimation of retention factors and spot separation.

**Separation Execution Device (SED).** SED integrates a BUCHI R300 rotary evaporator, multichannel solvent handling, an automated wet-loading module, and a SepaBean U100 flash-chromatography system. After solvent removal, the crude material is redissolved and transferred from the evaporation flask to the flash column by a syringe pump. An inline bubble sensor detects completion of liquid transfer and switches the flow path to divert residual air away from the silica column. SED then executes the programmed solvent gradient and returns the gradient and UV traces to SEA.

**Peptide Device (PD).** PD is the previously reported automated platform used for solid-phase peptide synthesis and on-resin head-to-tail cyclization. It comprises seven functional modules for system control, reagent storage, reagent delivery, flow-path switching, peptide coupling and cyclization, process detection, and product collection. PD executes the residue-specific coupling and cyclization workflow submitted by PA and returns the resulting product for HPLC–MS analysis.

## Code availability

Codes are publicly available in the GitHub repository (https://github.com/guikunchen/VibeSyn) and the Zenodo repository (https://zenodo.org/records/23764485).